\documentclass[prl,twocolumn,superscriptaddress,amsmath,amssymb,aps]{revtex4-1}
\usepackage{amsmath}
\usepackage{amssymb}
\usepackage{graphicx}
\usepackage{overpic}
\usepackage{siunitx}
\usepackage{hyperref}
\usepackage{color}
\usepackage{comment}

\DeclareMathOperator{\ku}{Ku}

\newcommand{\tauK}{\tau_{\eta}}

\newcommand\ve[1]{\boldsymbol{#1}}
\newcommand{\ma}[1]{\ensuremath{\mathbb{#1}}}
\newcommand{\vs}{v^{({\rm s})}}
\newcommand{\omegas}{\omega^{({\rm s})}}

\newcommand{\urms}{{u_{\rm rms}}}
\newcommand{\tr}{\ensuremath{\mbox{tr}}}
\newcommand{\trSS}{{{\mathcal S}^2}}
\newcommand{\trSSt}{{{\mathcal S}(t)^2}}
\newcommand{\Yc}{{{\nabla\mathcal S}^2_{\rm th}}}
\newcommand{\nhat}{{\hat{\ve n}}}
\newcommand{\phat}{{\hat{\ve p}}}
\newcommand{\qhat}{{\hat{\ve q}}}

\begin{document}

\title{Efficient survival strategy for zooplankton in turbulence}

\author{N. Mousavi}
\affiliation{Department of Physics, Gothenburg University, Gothenburg, SE-40530 Sweden}

\author{J. Qiu}
\affiliation{AML, Department of Engineering Mechanics, Tsinghua University, Beijing, 100084 China}

\author{B. Mehlig}
\affiliation{Department of Physics, Gothenburg University, Gothenburg, SE-40530 Sweden}

\author{L. Zhao}
\affiliation{AML, Department of Engineering Mechanics, Tsinghua University, Beijing, 100084 China}

\author{K. Gustavsson}
\affiliation{Department of Physics, Gothenburg University, Gothenburg, SE-40530 Sweden}

\date{\today}

\begin{abstract}
Zooplankton in a quiescent environment can detect predators by hydrodynamic sensing, triggering powerful escape responses. Since turbulent strain tends to mask the hydrodynamic signal, the organisms should avoid such regions, but it is not known how they accomplish this. We found a simple, robust, and highly efficient strategy, that relies on measuring the sign of gradients of squared strain. Plankton following this strategy show very strong spatial clustering, and align against the local flow velocity, facilitating mate finding and feeding. The strategy has  the potential to reconcile competing fitness pressures.
\end{abstract}

\keywords{Zooplankton, Turbulence, Counter-current swimming, Strainphobic}

\maketitle

\emph{Introduction.}--Zooplankton form an essential part of marine ecosystems, influencing both food webs and the climate~\cite{steinberg2017zooplankton}.
Their presence is vital for many fish species and aquaculture~\cite{stoettrup2000the,lomartire2021the}, their daily vertical migration influences the global carbon cycle~\cite{ariza2015migrant,steinberg2017zooplankton}, and they affect the albedo of the ocean~\cite{marcos2011microbial}.
Understanding plankton behavior is paramount for anticipating variations in their abundance, and this, in turn, requires insights into how plankton navigate within turbulent environments.

Simple models are tremendously successful in explaining behavior of swimming phytoplankton and bacteria in flows~\cite{pedley1992hydrodynamic,guasto2012fluid}.
For example, a model for gyrotactic microswimmers~\cite{kessler1985hydrodynamic} explains shear trapping in turbulence~\cite{durham2009disruption}, inhomogeneous spatial distribution~\cite{durham2013turbulence}, accumulation in down-welling~\cite{kessler1985hydrodynamic,durham2013turbulence} or up-welling regions~\cite{gustavsson2016preferential,lovecchio2019chain}.

Unlike phytoplankton, zooplankton use setae on their bodies and antennae~\cite{yen1992mechanoreception} to measure flow disturbances \cite{yen1990setal,fields2002mechanical}. This helps them detect and distinguish predators, mates, and food~\cite{pecseli2016plankton}.
Observations have revealed that many zooplankton species navigate efficiently in moderately turbulent flow, by adjusting their jump frequency and velocity~\cite{saiz1992free}, as well as their swimming pattern~\cite{michalec2015turbulence} in response to local flow characteristics, except under substantial turbulent fluctuations~\cite{quiniou2022copepod}.
In laminar flows, the strain rate triggers escape reactions.
Experiments demonstrate that in turbulent environments comparable or even larger magnitudes of strain may be ignored~\cite{adhikari2015simultaneous}.
Moreover, copepods can exert control over their turbulent diffusion by adjusting their jumping frequency, although the jumps are uncorrelated with the local strain rate and its history for at least two seconds~\cite{michalec2017zooplankton}.
Finally, copepods in small-scale vortices respond to vorticity rather than strain~\cite{elmi2020the,elmi2022copepod}, despite vorticity being hard to measure in their frame of reference.
These experimental observations indicate that information beyond the local strain rate matters for navigation in turbulence.
However, little is in general known about which signals and mechanisms are used for efficient navigation.

\begin{figure}[t]
    \begin{overpic}[width=0.494\textwidth]{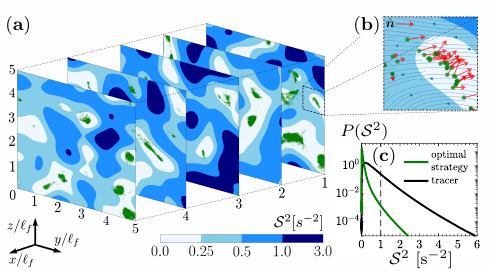}
    \end{overpic}
    \caption{
        ({\bf a}) Snapshot showing five spatial slices with positions of swimmers (green points) following the optimal strategy (\ref{eq:optimal_policy}) in the stochastic model (see text). Flow strain $\trSS$ is color-coded.
        ({\bf b}) Zoom, including flow streamlines and swimmer direction projected onto the image plane.
        ({\bf c}) Steady-state probability distributions of $\trSS$ evaluated along swimmer trajectories (green) and for tracer particles (black).
        Parameters $\lambda=2$, $\vs=\SI{20}{\milli\meter\per\s}$, $\omegas=\SI{5}{\radian\per\s}$, $\nu=\SI{1}{\milli\meter\squared\per\s}$ and $\varepsilon=\SI{1}{\milli\meter\squared\per\s\cubed}$ ($\urms=\SI{10}{\milli\meter\per\s}$ and $\tauK=\SI{1}{\s}$).
    }
    \label{fig:positions}
\end{figure}

The perhaps most critical task for zooplankton is evading predators.
Successful detection of the flow disturbance ahead of an approaching predator swimming faster than flow-dispersed chemical signals, offers an opportunity to escape~\cite{kiorboe2008mechanistic}.
In laminar flow, escapes are triggered by the magnitude of the strain rate tensor being above a critical threshold, varying by species from \SIrange{0.4}{6}{\per\s}~\cite{buskey2002escape,kiorboe1999hydrodynamic,jakobsen2001escape}.
Turbulent strain is harmful because it impedes escape by masking the predator signal~\cite{pecseli2016plankton}. Additionally, it can trigger false alarms, putting the zooplankton at risk by revealing its location when jumping in response to such signals~\cite{yen1996advertisement}.
Efficient predator evasion, therefore, requires the plankton to find and remain in low-strain regions.
Certain species migrate to calmer layers when faced with regions of high mean shear or turbulence intensity, and thus, high strain regions~\cite{ianson2011response,incze2001changes,visser2009swimming}, while others stay, perhaps to enhance prey contact~\cite{visser2009swimming}.
However, it is not known how zooplankton succeed in avoiding high-strain regions in turbulent flow.
Which are the most important hydromechanical signals, and which are the most efficient strategies to achieve this goal?
Unveiling such strategies offers insight into the intricate interplay between zooplankton behavior, predator evasion, and turbulent flow dynamics.
In order to answer these questions, we formulate a model for cruising zooplankton that can actively adapt swimming speed and orientation, in response to flow signals.
We identify key signals, and determine the optimal strategy that allows cruising zooplankton to efficiently avoid high-strain regions across a wide range of turbulent dissipation rates.
Fig.~\ref{fig:positions}({\bf a}) illustrates the remarkable success of this strategy.

Related navigation problems, where microswimmers in complex flow target either an absolute point or direction, have recently been addressed using analytical approaches~\cite{liebchen2019optimal,biferale2019zermelo,moussaIder2021hydrodynamics,monthiller2022surfing,piro2023energetic} or reinforcement learning~\cite{colabrese2017flow,biferale2019zermelo,schneider2019optimal,alageshan2020machine,gunnarson2021learning,muinos2021reinforcement,qiu2022navigation,qiu2022active,xu2023long}.
Here, we investigate for the first time the vital task of how zooplankton can avoid local high-strain regions fluctuating in space and time in turbulent flows, using realistic sensing abilities and characteristics of the swimmer.
In an earlier model for copepod clustering~\cite{ardeshiri2016lagrangian}, copepods jump upon encountering strain rates above a set value.
Although reducing time spent in high-strain zones, this strategy is not optimal because the copepod moves through the high-strain areas. Our novel strategy ensures zooplankton avoid high-strain regions altogether.

\emph{Model.}--Refs.~\cite{saiz1992free,adhikari2015simultaneous,michalec2015turbulence,michalec2017zooplankton,elmi2020the,elmi2022copepod,quiniou2022copepod} illustrate that zooplankton do respond to stationary and turbulent flow. The problem is that very little is known about which the most important signals are and how these are used for navigation. We therefore ask the question how zooplankton {\em should} behave in a flow to solve the vital task of avoiding high-strain regions masking the signal from approaching predators.
The aim of our model is to capture the essential dynamics of cruising zooplankton, allowing to explore mechanisms applicable to a broad range of species.
We consider a single organism and analyse how it can navigate in response to hydromechanical signals.
Assuming the zooplankton can swim, steer and sense its surroundings, its velocity
\begin{subequations}
\label{eq:eom}
\begin{align}
\ve v(t)=\ve u(\ve x,t)+\vs(t)\nhat(t)
\end{align}
is composed of the flow $\ve u$ and active swimming with speed $\vs(t)$ in its direction $\nhat$.
This model is widely used to discuss all kinds of microorganisms swimming in the ocean~\cite{kessler1985hydrodynamic,durham2009disruption,durham2013turbulence,gustavsson2016preferential,lovecchio2019chain,monthiller2022surfing,piro2023energetic,colabrese2017flow,alageshan2020machine,gunnarson2021learning}.
Angular velocity $\ve\omega$ is influenced by fluid vorticity $2\ve\Omega=\ve\nabla\times\ve u$, strain rate tensor $\ma S$ with components $S_{ij}=(\partial_iu_j+\partial_ju_i)/2$, and active steering, $\omegas_p(t)$ and $\omegas_q(t)$, around orthogonal axes $\phat$ and $\qhat$ transversal to $\nhat$ (steering around $\ve n$ does not matter due to symmetries).
Approximating the shape by a spheroid with aspect ratio $\lambda$, the dynamics becomes~\cite{jeffery1922motion},
\begin{align}
\ve\omega=\ve \Omega(\ve x,t) + \frac{\lambda^2-1}{\lambda^2+1} \nhat \times \mathbb{S}(\ve x , t) \nhat + \omegas_p(t)\phat + \omegas_q(t)\qhat\,.
\end{align}
\end{subequations}
The dynamics is controlled by choosing $\vs(t)$, $\omegas_p(t)$ and $\omegas_q(t)$ within maximal speed $\vs$ and angular speed~$\omegas$.

Zooplankton that jump in response to strain signals to escape predators vary greatly in size~\cite{buskey2002escape,kiorboe1999hydrodynamic,jakobsen2001escape}, from ciliates as small as $\SI{20}{\micro\meter}$ to copepods with length $L$ from $\SIrange{0.6}{5}{\milli\meter}$~\cite{svetlichny2020kinematic}.
Since larger zooplankton in general swim faster, we expect them to avoid high-strain regions more easily.
Swimming speeds of copepods vary from $\SIrange{1}{50}{\milli\meter\per\s}$~\cite{svetlichny2020kinematic}.
Our model assumes a slighlty prolate shape ($\lambda=2$), length $L=\SI{2}{\milli\meter}$, and swimming speed $\vs=\SI{20}{\milli\meter\per\s}$.
Cruising copepods execute rapid turns~\cite{andersen2012kinematics, morris1990mechanics}, but references for typical values of $\omega(s)$ are lacking.
We conservatively use $\omegas=\SI{5}{\radian\per\s}$, one sixth of the diameter-to-swimming-speed ratio estimated in Refs.~\cite{qiu2022active,monthiller2022surfing}.
In turbulence of moderate intensity, the precise value of $\omegas$ does not matter much for our findings, while a larger value of $\omegas$ is advantageous if the turbulence intensity is very high.

In our simulations, we employ homogeneous and isotropic turbulence, either from direct numerical simulations (DNS)~\cite{qiu2022active,JohnsHopkins,JohnsHopkins2,qiu2019settling} or a stochastic model~\cite{gustavsson2016statistical,bec2024statistical}.
The latter has a single spatial scale $\ell_{\rm f}$. Exponential tails of fluid gradients model non-Gaussian gradients in turbulence.
Dynamics of swimmers or inertial particles in this model qualitatively, and in many cases quantitatively, matches results from DNS~\cite{gustavsson2016statistical,bec2024statistical}.
See Appendix~\ref{sec:simulations} for details on the flows.

\emph{Optimal strategy.}--The aim is to determine the most important flow signals and how to exploit them to avoid high-strain regions. To this end, we minimize the squared strain rate, $\trSSt=\tr(\ma S(\ve x(t),t)^2)$, along the trajectory of a swimmer following the active part of its dynamics, by making an expansion in a short time interval $\delta t$ to lowest contributing order in $\delta t\,\vs(t)$, $\delta t\,\omegas_p(t)$, and $\delta t\,\omegas_q(t)$
\begin{align}
    \label{eq:TrSSqr_org}
     & {\mathcal S}(t+\delta t)^2={\cal S}(t)^2+
    \delta t[\partial_t\trSSt
            +\vs(t)\nhat(t)\cdot\ve\nabla\trSSt]
    \nonumber                                    \\&
    \hspace{0.4cm}+\frac{1}{2}\delta t^2\vs(t)[\omegas_q(t)\phat(t)-\omegas_p(t)\qhat(t)]\cdot\ve\nabla\trSSt\,.
\end{align}

The term linear in $\delta t$ is minimized by the control
\begin{subequations}
    \label{eq:optimal_policy}
    \begin{align}
        \label{eq:optimal_propulsion}
        \vs_{\rm opt}(t) & =\left\{
        \begin{array}{ll}
            \vs & \mbox{if }\nhat \cdot \ve \nabla \trSSt<0\cr
            0   & \mbox{otherwise}
        \end{array}\right.\,,
    \end{align}
    and the $\delta t^2$-term is minimized by the control
    \begin{align}
        \label{eq:optimal_rotationp}
        \omegas_{p,{\rm opt}}(t) & =\omegas\text{sign}(\qhat \cdot \ve \nabla \trSSt)\,, \\
        \omegas_{q,{\rm opt}}(t) & =-\omegas \text{sign}(\phat\cdot\ve\nabla\trSSt)\,.
        \label{eq:optimal_rotationq}
    \end{align}
\end{subequations}
We find that the most important signal for short-term strain avoidance are the signs of projections of the squared strain gradients on the swimming direction $\nhat$, and the transversal directions $\phat$ and $\qhat$.
The resulting strategy is to swim if the strain decreases in the swimming direction, and steer such that $\ve n$ rotates towards the direction of steepest strain descent.
We have performed reinforcement learning~\cite{sutton2018reinforcement}, with resulting strategies indicating that Eq.~(\ref{eq:optimal_policy}) is also optimal in the long run [unpublished].

By superposing swimmers starting from different initial conditions, Fig.~\ref{fig:positions}({\bf a}) illustrates the spatial probability of a single swimmer following the optimal strategy (\ref{eq:optimal_policy}) in the stochastic model with moderate turbulence.
The swimmer avoids high-strain regions, as desired, and accumulates in regions of low strain.
Fig.~\ref{fig:positions}({\bf c}) shows the probability of strain along its trajectory (green line).
Strain exceeding the threshold for predator detection (\SI{\sim 1}{\per\s\squared}, dashed line) is far rarer than for tracers (black line).

\begin{figure}[t]
    \begin{overpic}[width=0.48\textwidth]{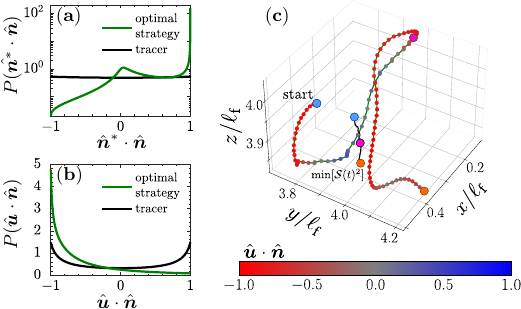}
    \end{overpic}
    \caption{
        Probability distributions of ({\bf a}) $\nhat^*\cdot\nhat$ and ({\bf b}) $\hat{\ve u}\cdot\nhat$ for the parameters in Fig.~\ref{fig:positions}.
        ({\bf c}) Example trajectory (green) with color coded $\hat{\ve u}\cdot\nhat$ (small markers) for duration $\sqrt{5}\tauK$. Large markers denote start (blue), midpoint (pink), and end (orange). The black trajectory shows the location of the only local strain minimum in the displayed region.
    }
    \label{fig:alignment}
\end{figure}

The mechanism for avoiding high-strain regions can be understood as follows.
First, a swimmer following Eq.~(\ref{eq:optimal_propulsion}) decreases its instantaneous value of $\trSS$ by only swimming if the strain decreases ahead of it.
Second, the angular dynamics in Eqs.~(\ref{eq:optimal_rotationp},\ref{eq:optimal_rotationq}) has fixed points when both $\phat$ and $\qhat$ are perpendicular to $\ve{\nabla}\trSS$, i.e. when the swimming direction $\nhat$ is equal to either of $\pm \ve\nabla\trSS/\lvert\ve\nabla\trSS\rvert$.
The stable fixed point $\nhat^*=-\ve\nabla\trSS/\lvert\ve\nabla\trSS\rvert$ is the direction in which $\trSS$ decreases most.
Due to the turbulent fluctuations, $\nhat$ does not follow $\nhat^*$ perfectly.
Fig.~\ref{fig:alignment}({\bf a}) shows the distribution of $\nhat^*\cdot\nhat$.
There is a strong bias to align with $\nhat^*$: $80\%$ of the swimmers have positive $\nhat^*\cdot\nhat$. But the alignment is not perfect, only $25\%$ have $\nhat^*\cdot\nhat>0.8$, belonging to the sharp peak at unity.
Nevertheless, the alignment bias allows the swimmers to efficiently avoid high strain.

The swimmers in Fig.~\ref{fig:positions}({\bf a}) accumulate in low-strain regions.
This constrains the phase-space dynamics compared to tracer particles, meaning that unexpected correlations may emerge.
One example is a tendency for the swimmers to swim against the streamlines of the flow, see Fig.~\ref{fig:positions}({\bf b}). Fig.~\ref{fig:alignment}({\bf b}) shows the distribution of the alignment between the direction of the swimmer, $\nhat$, and the flow, $\hat{\ve u}$.
The distribution is strongly skewed towards anti-alignment, with an average $\langle\hat{\ve u}\cdot\nhat\rangle\approx-0.5$.
This may be perceived as surprising: in models of phytoplankton, where $\vs$ is constant and $\omegas=0$, the swimmer instead has a bias to align with the flow~\cite{borgnino2019alignment,borgnino2022alignment}.

To explain the alignment against the flow, we consider a simplified model in the limit $\omegas\to\infty$. Then swimmers quickly align with the stable orientation $\nhat^*$.
The strategy (\ref{eq:optimal_policy}) approaches gradient descent of $\trSS$ subject to advection by the flow. The velocity simplifies to
\begin{align}
    \label{eq:eqm_simplified}
    \ve v=\ve u(\ve x,t)-\vs \ve\nabla\trSS/\lvert\ve\nabla\trSS\rvert\,.
\end{align}
This effective velocity field has non-zero divergence, showing that particles do not distribute uniformly in space.
A flow speed much larger than the swimming speed hinders efficient strain avoidance.
In the opposite limit, swimmers exactly follow local minima of the strain.
When both speeds are of the same order, the swimmer circulates closely around the strain minimum, but fails to reach it because of the flow.
Since the velocity is larger when swimming with the flow than against it, the swimmer spends more time in flow regions where it swims against the flow. This effect becomes more prominent the closer the swimming speed is to the flow speed.
The same mechanisms applies for finite $\omega$.
One example trajectory is shown in Fig.~\ref{fig:alignment}({\bf c}).
The velocity is given by the displacement between successive markers, color coded by the alignment with the flow.
The trajectory encircles the strain minimum, moving slower when anti-aligned with the flow velocity.
This explains the alignment against the flow in Fig.~\ref{fig:alignment}({\bf b}).
We remark that the found mechanism is kinematic and unrelated to the task of minimizing $\trSS$. It gives rise to counter-current swimming for swimmers tracking a generic point target in the presence of turbulent fluctuations.

The results discussed so far assumes that the signs of the gradients of squared strain in Eq.~(\ref{eq:optimal_policy}) can be measured perfectly.
However, mechanoreceptive zooplankton have a finite resolution in their measurements.
To investigate when they are able to follow the strategy, we estimate a sensitivity threshold $\Yc$ of components of $\ve\nabla\trSS$.
The setae of copepods can measure velocity differences between their body and the ambient fluid down to $\Delta u_{\rm th}\sim\SI{10}{\micro\meter\per\s}$~\cite{yen1992mechanoreception,jiang2008hydrodynamic}.
An estimate of the lower limit of $\Yc$ is $\Delta u_{\rm th}^2/(L/2)^3\sim\SI{0.1}{\per\meter\per\s\squared}$, for a swimmer of length $L=\SI{2}{\milli\meter}$.
Since zooplankton are hardly able to measure all components with this precision while cruising, we use a ten times larger threshold in our simulations, $\Yc=\SI{1}{\per\meter\per\s\squared}$.
When the magnitude of components of $\ve\nabla\trSS$ are below this threshold, we set $\omegas(t)$ to 0, and set $\vs(t)$ randomly to either $\vs$ or $0$, keeping this value until the signal becomes larger than the threshold.
Simulations for moderate turbulence intensity show that the exact value of $\Yc$ is not important (see Fig.~\ref{fig:distributions}({\bf a}) in Appendix~\ref{sec:robustness}).
The distribution with $\Yc=\SI{1}{\per\meter\per\s\squared}$ is identical to that obtained without a threshold.
Even a 100 times larger threshold gives the same distribution.
Larger thresholds reduce the performance, but even for a threshold as large as \SI{1000}{\per\meter\per\s\squared}, there is a clear reduction in probability to sample very large strain.
In conclusion, the exact value of the threshold does not matter for intermediate turbulent intensities.
For larger turbulence intensities, $|\ve\nabla\trSS|$ is typically larger, making the dynamics even less sensitive to the threshold.
For smaller turbulence intensities, $|\ve\nabla\trSS|$ is smaller, making the value of the threshold more important. However, for that case, strain is anyway small, meaning that it is not as important to be able to minimize it.

In the ocean, the energy dissipation rate per unit mass, $\varepsilon$, spans from $10^{-4}\,\SI{}{\milli\meter\squared\per\s\cubed}$ in the deep ocean to $\SI{100}{\milli\meter\squared\per\s\cubed}$ in the upper ocean layer~\cite{yamazaki1996comparison,fuchs2016seascape}.
The root-mean squared velocity $\urms=\langle\ve u^2\rangle^{1/2}$ ranges from \SIrange{0.1}{100}{\milli\meter\per\s}~\cite{yamazaki1996comparison}.
In our simulations, we use $\urms=10\sqrt{\varepsilon}\;\SI{}{\s\tothe{1/2}}$ fitted to the data in Ref.~\cite{yamazaki1996comparison}, and a kinematic viscosity $\nu=\SI{1}{\milli\meter\squared\per\s}$.
The latter amounts to Kolmogorov times $\tauK$ from \SI{100}{\s} down to \SI{0.1}{\s}, Kolmogorov lengths $\eta$ from \SI{10}{\milli\metre} to \SI{0.3}{\milli\metre}, and Taylor-scale Reynolds numbers from $1$ to $1000$.
Typical response times of copepods are a few milliseconds~\cite{lenz1999reaction,buskey2002escape,lenz2004force}, well below the Kolmogorov time.
The zooplankton length, $L=\SI{2}{\milli\metre}$, is below the smooth length scale, $\sim 10\eta$, where the dissipation and inertial ranges cross~\cite{pecseli2016plankton,bec2024statistical}.

\begin{figure}[t]
    \includegraphics[width=7.5cm]{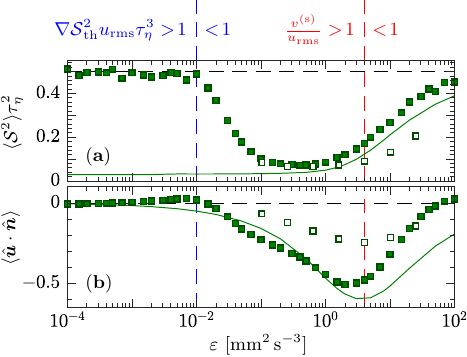}
    \caption{
        \label{fig:sweep}
        Strain avoidance and counter-current alignment against the turbulent dissipation rate $\varepsilon$.
        ({\bf a}) Average strain $\langle\trSS\rangle$ in the stochastic model (filled markers) and DNS with ${\rm Re}_\lambda\approx 60$ (empty markers).
        Solid line shows model results for the simplified dynamics (\ref{eq:eqm_simplified}).
        Horizontal dashed line shows value for tracer particles.
        Vertical dashed lines show where $\vs/\urms=1$ and $\Yc\urms\tauK^3=1$.
        The dimensionless rotational swimming speed is $\omegas\tauK=2.5\vs/\urms$.
        ({\bf b}) Same as panel ({\bf a}), but for the average counter-current alignment, $\langle\hat{\ve u}\cdot\nhat\rangle$.
        Parameters $\lambda=2$, $\vs=\SI{20}{\milli\meter\per\s}$, $\omegas=\SI{5}{\radian\per\s}$, $\Yc=\SI{1}{\per\meter\per\s\squared}$ and $\nu=\SI{1}{\milli\meter\squared\per\s}$.
    }
\end{figure}

Fig.~\ref{fig:sweep} shows averages of $\trSS\tauK^2$ ({\bf a}) and $\hat{\ve u}\cdot\ve n$ ({\bf b}) against $\varepsilon$.
First, solid lines show stochastic-model results for the simplified model (\ref{eq:eqm_simplified}). In accordance with the analysis below Eq.~(\ref{eq:eqm_simplified}), the average strain in panel ({\bf a}) is exceptionally low due to tracking of strain minima when $\vs/\urms\gg 1$ (to the left of the vertical red dashed line), becomes slightly higher due to circulation around minima when $\vs/\urms\sim 1$, and approaches the level of tracer particles when $\vs/\urms\ll 1$.
The alignment against the flow in panel ({\bf b}) is strong when $\vs\sim \urms$, consistent with the mechanism for counter-current swimming described above.
In contrast, for the case where $\vs/\urms\gg 1$, the flow is too weak to give a substantial effect and when $\vs/\urms\ll 1$  the alignment is small because the swimmer fails to track the strain minima.

Second, filled markers in Fig.~\ref{fig:sweep} show stochastic-model results for swimmers following the optimal strategy~(\ref{eq:optimal_policy}) with a sensing threshold.
The main differences to the simplified model are that $\langle\trSS\rangle$ is much larger for the smallest $\varepsilon$ and that both $\langle\trSS\rangle$ and $\langle\hat{\ve u}\cdot\ve n\rangle$ are slightly larger for the largest $\varepsilon$.
Simulations with the threshold set to zero (Fig.~\ref{fig:sweepSupplemental} in Appendix~\ref{sec:sweep_no_threshold}) reveal that the first difference is entirely due to the sensing threshold, which becomes important for $\Yc$ larger than its characteristic scale $1/(\urms\tau_\eta^3)$ (vertical blue dashed line). The second difference entirely arises from the angular velocity being finite, slightly reducing the performance compared to the simplified model when flow gradients are vigorous.

Finally, results for the optimal strategy from DNS are plotted as empty markers.
Both $\langle\trSS\rangle$ and $\langle\hat{\ve u}\cdot\ve n\rangle$ show similar trends as the stochastic model, with minima around $\vs=\urms$.
The degree of preferential sampling of strain is of the same order, but the degree of alignment is weaker in the DNS.
The latter is expected because the DNS involves a range of scales, in contrast to the single velocity scale in the stochastic model. As a result, while the swimmer circulates around the strain minimum, the chance of encountering an opposing flow velocity that matches its swimming speed decreases.

\begin{figure*}[t]
    \begin{overpic}[width=\textwidth]{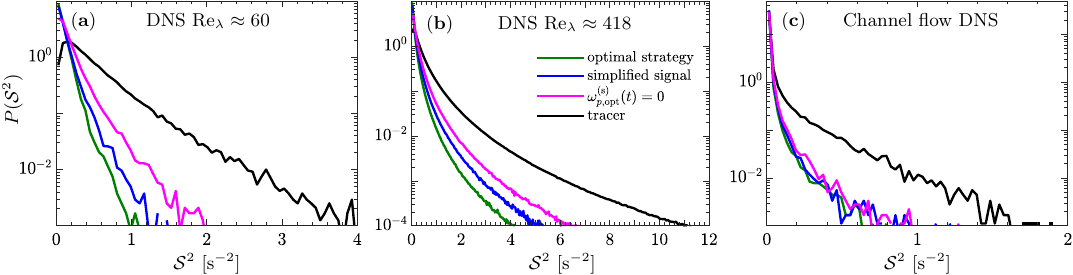}
    \end{overpic}
    \caption{
        \label{fig:distributionsFlows}
        Strain rate distributions for the optimal strategy~(\ref{eq:optimal_policy}) (green), the simplified signal $\tilde{\mathcal S}$ (blue), the simplified signal with $\omegas_{p,{\rm opt}}(t)$ put to zero (magenta), and tracer particles (black) in DNS of homogeneous isotropic turbulence with ({\bf a}) ${\rm Re}_\lambda\approx 60$, and ({\bf b}) ${\rm Re}_\lambda\approx 418$, and channel flow turbulence with  friction Reynolds number ${\rm Re}_\tau=180$ ({\bf c}).
        Parameters as in Fig.~\ref{fig:positions}({\bf c}).
    }
\end{figure*}

Fig.~\ref{fig:sweep} shows that the optimal strategy (\ref{eq:optimal_policy}) with sensing thresholds is very efficient for avoiding high-strain regions for a large range of dissipation rates $\varepsilon$ that are not too extreme.
To explore the robustness of this strategy to the variety of flows encountered in the ocean, we have confirmed its effectiveness at different Reynolds numbers and with a mean flow profile.
Fig.~\ref{fig:distributionsFlows} shows strain-rate distributions akin to Fig.~\ref{fig:positions}({\bf c}) in DNS of turbulence with dissipation rate $\varepsilon=\SI{1}{\milli\metre\squared\per\s\cubed}$ (see Appendix~\ref{sec:simulations} for details on the flows).
Panels ({\bf a},{\bf b}) present results from DNS of homogeneous isotropic turbulence with Taylor-scale Reynolds number ({\bf a}) ${\rm Re}_\lambda\approx 60$ and ({\bf b}) ${\rm Re}_\lambda\approx 418$, the latter integrated using data from Johns Hopkins University turbulence database~\cite{JohnsHopkins,JohnsHopkins2}.
We find that large strain gradients have orders of magnitude lower probability compared to tracer particles, in qualitative agreement with the stochastic model in Fig.~\ref{fig:positions}({\bf c}).
The reported Reynolds number ${\rm Re}_\lambda\approx 130$ of ocean turbulence with $\varepsilon=\SI{1}{\milli\metre\squared\per\s\cubed}$~\cite{yamazaki1996comparison} lies in between, indicating that the strategy is relevant for turbulent ocean conditions.
Additionally, Fig.~\ref{fig:distributionsFlows}({\bf c}) displays results from DNS of a turbulent channel flow~\cite{qiu2019settling}, demonstrating the effectiveness of the strategy in non-homogeneous flows typical of ocean turbulence.
In conclusion, Figs.~\ref{fig:sweep}, \ref{fig:distributionsFlows}, and \ref{fig:distributions} show that the strategy is robust, performing well in a wide range of flow configurations and swimmer parameters.

\emph{Discussion.}--Our analysis shows that the sign of gradients of squared strain are the most important signals for avoiding high-strain regions.
Gradients are important for optimizing more general scalar quantities in microswimmer dynamics, for example chemotaxis~\cite{macnab1972the}, light intensity~\cite{jekely2009evolution}, point-to point navigation~\cite{zermelo1931uber,biferale2019zermelo}, or vertical navigation~\cite{monthiller2022surfing,piro2023energetic}.
However, the dynamics and strategy taken in response to the gradients are different in each example.
We remark that chemotaxis can also be obtained by sampling the history of a concentration signal~\cite{segall1986temporal}.
In our analysis we did not include history, but it is expected that the history of strain is correlated to the strain gradients, meaning that measuring history, or other quantities that are correlated to the strain gradients, may be preferred for species unable to measure strain gradients.
As an illustration, assume that the swimmer uses the bending pattern of its setae to measure spatial variations of the maximal strain component $\tilde{\mathcal S}(\ve x,t)={\rm max}_{i,j}|S_{ij}(\ve x,t)|$ in its local frame.
This signal is simpler than $\mathcal S$ because it only depends on the maximal strain component and it is linear.
Results obtained by replacing $\ve\nabla\mathcal S$ in Eq.~(\ref{eq:optimal_policy}) by $\ve\nabla\tilde{\mathcal S}$, shows that the performance is almost as good for the simplified signal, see blue curves in Fig.~\ref{fig:distributionsFlows}.
Additionally, assuming that the swimmer only performs rotational swimming around the $\ve q$-axis, i.e. $\omegas_{p,{\rm opt}}(t)=0$, it only needs to measure the signs of $\nhat\cdot\ve\nabla\tilde{\mathcal S}$ and $\phat\cdot\ve\nabla\tilde{\mathcal S}$ in Eq.~(\ref{eq:optimal_policy}). These are obtained by measuring $\tilde{\mathcal S}$ at different locations in the $\nhat$-$\phat$ (tail-antannae) plane, where it has highest resolution.
Fig.~\ref{fig:distributionsFlows} (magenta) shows that the strategy is slightly worse, but still significantly better than tracer particles.

In regions far below the strain threshold level, it could be beneficial to change strategy to achieve other goals.
However, supposing the zooplankton continues to follow the strategy~(\ref{eq:optimal_policy}), they accumulate close to strain minima, which may be beneficial for mate finding in the vast ocean.
Moreover, as they circulate around the strain minimum, they exhibit counter-current swimming if either their swimming speed lies close to that of the flow, or if they adjust it to that of the flow.
From simple geometrical arguments, counter-current swimming increases the rate of head-on encounters with prey or food particles advected by the flow, and may facilitate detection of chemical cues~\cite{chapman2011animal}.
Consequently, the strategy we have identified offers a potential solution to the significant challenge of balancing competing fitness pressures, such as maintaining a high feeding rate for growth and reproduction while simultaneously minimizing the risk of predation~\cite{visser2009swimming}.

Ocean flows are often stratified with anisotropic large-scale structures.
In our DNS of a turbulent channel flow, we found that the strategy (\ref{eq:optimal_policy}) performs well in an anisotropic shear flow.
Fig.~\ref{fig:distributionsFlows}({\bf c}) shows that swimmers avoid high-strain regions and accumulate near the center where strain is minimal.
More generally, we expect that this strategy helps avoid long-lived large-scale structures in the ocean.
While occasional formation of short-lived local strain minima may pose temporary obstacles, large-scale high-strain regions are avoided in the long run if their strain gradients lie below the sensing threshold, or if they exhibit a preference for larger strain fluctuations compared to low-strain regions.
It remains an open question whether this mechanism contributes to the migration from large-scale turbulent regions~\cite{ianson2011response,incze2001changes,visser2009swimming}, or for counter-current swimming against large-scale currents observed in various zooplankton species~\cite{genin2005swimming,shang2008resisting,sidler2018counter}.

Because of the tremendous challenge to acquire simultaneous experimental data for both swimmers and flow, the extent to which zooplankton actively avoid turbulent strain at small scales remains an unanswered question.
However, recent development of imaging technologies such as high-speed cameras~\cite{sidler2017three}, acoustic imaging~\cite{ohman2018zooglider}, and holographic microscopy~\cite{bachimanchi2022microplankton}, raises hope to answer this question in the near future.

\emph{Acknowledgements.}--KG was supported by Vetenskapsr\aa{}det (grant no.~2018-03974).
BM was supported by Vetenskapsr\aa{}det (grant no.~2021-4452).
LZ and JQ are grateful for the support of the Natural Science Foundation of China (grants nos. 92252104 and 91752205).
The statistical-model simulations were conducted using the resources provided by the Swedish National Infrastructure for Computing (SNIC).

\setcounter{secnumdepth}{3}

\renewcommand{\thesection}{\Roman{section}}
\renewcommand{\thesubsection}{\Roman{section}.\arabic{subsection}}

\appendix

\section{Numerical simulations}
\label{sec:simulations}
Below, we outline our simulations of various flow setups: the stochastic turbulence model (Sec.~\ref{sec:stochastic}, Figs.~1--3, 5 and 6), DNS of homogeneous isotropic turbulence with ${\rm Re}_\lambda\approx 60$ (Sec.~\ref{sec:DNSsmall}, Figs.~\ref{fig:sweep} and \ref{fig:distributionsFlows}), DNS of homogeneous isotropic turbulence with ${\rm Re}_\lambda\approx 418$ (Sec.~\ref{sec:DNSlarge}, Fig.~\ref{fig:distributionsFlows}), and turbulent channel flow (Sec.~\ref{sec:DNSchannel}, Fig.~\ref{fig:distributionsFlows}).

\subsection{Stochastic flow model}
\label{sec:stochastic}
We model turbulence using an incompressible random flow with homogeneous and isotropic statistics in three spatial dimensions. It is a non-Gaussian generalisation of the Gaussian random flow reviewed in Refs.~\cite{gustavsson2016statistical,bec2024statistical}.
The flow $\ve u(\ve x,t)$ is generated as a superposition of $M$ time-independent random velocity fields $\ve u_m(\ve x)$
\begin{align}
    \ve u(\ve x,t)=\sum_{m=1}^Mc_m(t)\ve u_m(\ve x)\,.
    \label{eq:u_superposition}
\end{align}
The coefficients $c_m(t)$ follow independent Ornstein-Uhlenbeck processes with time scale $\tau_{\rm f}$ and variance $M^{-1}$, i.e. they have zero mean and correlation functions
\begin{align}
    \langle c_m(t)c_{n}(t')\rangle=\frac{1}{M}\delta_{mn}e^{|t-t'|/\tau_{\rm f}}\,.
\end{align}
The fields $\ve u_m(\ve x)$ are generated through independent spatially smooth Gaussian random vector potentials~\cite{gustavsson2016statistical,bec2024statistical}, defined as $\ve u_m(\ve x)=\ve\nabla\times\ve\Psi_m(\ve x)$. The components $\Psi_{m,i}$ have zero mean and correlation functions
\begin{align}
    \langle \Psi_{m,i}(\ve x ) \Psi_{n,j}(\ve x') \rangle = \frac{1}{6}\delta_{mn}\delta_{ij}u_{\rm f}^2 \ell_{\rm f}^2 e^{-\lvert \ve{x}' - \ve{x} \rvert ^2/(2 \ell_{\rm f}^2)}\,.
\end{align}
Here $u_{\rm f}=\urms=\langle\ve u^2\rangle^{1/2}$ is the root-mean-squared velocity, $\ell_{\rm f}$ is the characteristic length-scale, and $\tau_{\rm f}$ is the correlation-time of the flow.

If the flow decorrelates faster due to spatial displacements than due to the Eulerian time scale $\tau_{\rm f}$, the dynamics of swimmers or inertial particles agree well with results from DNS~\cite{gustavsson2016statistical,bec2024statistical}.
This corresponds to large Kubo numbers, $\ku = \tau_{\rm f} u_{\rm f}/\ell_{\rm f} \gg 1$, where $\ell_{\rm f}/u_{\rm f}$ is proportional to the Lagrangian correlation time of the flow. This correlation time is represented by the Kolmogorov time $\tauK=\langle\tr(\ma A^{\rm T}\ma A)\rangle^{-1/2}=\ell_{\rm f}/(\sqrt{5}u_{\rm f})$, where the fluid gradient matrix $\ma A$ has components $A_{ij}=\partial_ju_i$.
In our simulations, we use $\ku=10$ in keeping with this limit.

Even though individual velocity fields $\ve u_m(\ve x)$ are Gaussian distributed, the superposition (\ref{eq:u_superposition}) is non-Gaussian if $M$ is finite.
The distributions of individual components of $\ve u$ and $\ma A$ have Gaussian body and exponential tails.
In our simulations, we use a finite value, $M=10$, to model non-Gaussian tails of fluid gradients.

\begin{figure*}[t]
    \begin{overpic}[width=\textwidth]{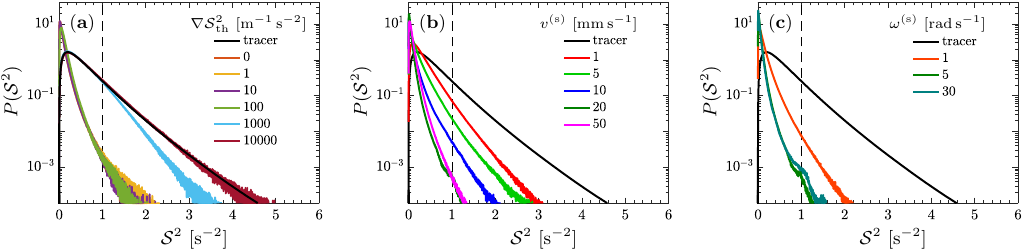}
    \end{overpic}
    \caption{
        \label{fig:distributions}
        Distribution of $\trSS$ in the turbulence model for ({\bf a}) different levels of the sensing threshold $\Yc$, ({\bf b}) swimming speed $\vs$, and ({\bf c}) angular swimming speed $\omegas$.
        Parameters $\lambda=2$, $\nu=\SI{1}{\milli\meter\squared\per\s}$ and $\varepsilon=\SI{0.64}{\milli\meter\squared\per\s\cubed}$.
        Unless otherwise stated, $\vs=\SI{20}{\milli\meter\per\s}$, $\omegas=\SI{5}{\radian\per\s}$, $\Yc=0$.
    }
\end{figure*}

\subsection{Homogeneous isotropic turbulence, ${\rm Re}_\lambda\approx 60$}
\label{sec:DNSsmall}
We use in house code~\cite{qiu2022active} for DNS of a homogeneous isotropic turbulent flow $\ve u$, using a pseudo-spectral method to solve the Navier-Stokes equations,
\begin{subequations}
\label{eq:NSapp}
\begin{align}
    \frac{\partial \ve{u}}{\partial t} + \ve{u}\cdot \nabla \ve{u} &= -\frac{1}{\rho_f}\nabla p_{\rm f}
    + \nu\nabla^2\ve{u} + \ve{f},
    \label{eq:NS1app}\\
    \nabla \cdot \ve{u} &= 0,
    \label{eq:NS2app}
\end{align}
\end{subequations}
where $p_{\rm f}$, $\rho_{\rm f}$ and $\nu$ are the pressure, density and the kinematic viscosity of the fluid.
To sustain turbulence by balancing viscous dissipation, we apply an external force~$\ve{f}$ at large scales~\cite{Machiels1997}.
We use periodic boundary conditions on a cubic domain with size $(2\pi)^3$.

The Taylor-scale Reynolds number, defined as $\mathrm{Re}_\lambda=u_{\rm rms}L_\lambda /(\sqrt{3}\nu)$, is set to $\mathrm{Re}_\lambda \approx 60$, with Taylor length scale $L_\lambda = u_{\rm rms} \sqrt{5\nu\varepsilon^{-1}}$. In order to resolve the velocity at the dissipation scales, we use $96$ grid points in each dimension. The smallest resolved scale is 1.78 times smaller than the Kolmogorov length scale, which means that the finest turbulent motion can be resolved~\cite{pope2000turbulent}.
The initial flow is random with exponential energy spectrum. We use second order Adams-Bashforth scheme for the time advancement of Eqs.~(\ref{eq:NSapp}) with a time step approximately ten times smaller than the Kolmogorov time scale.

Once the turbulence becomes fully developed, swimmers are initialized with random positions and orientations. A second-order Adams-Bashforth scheme is used to evolve the dynamics of the swimmers in Eqs.~(\ref{eq:eom}) with translational and rotational swimming velocities according to Eqs.~(\ref{eq:optimal_policy}).
The fluid velocity and its gradients at the swimmer position are interpolated using a second-order Lagrangian interpolation method from the values at Eulerian grid points.

\subsection{Homogeneous isotropic turbulence, ${\rm Re}_\lambda\approx 418$}
\label{sec:DNSlarge}
We used a DNS of forced homogeneous isotropic turbulence on a $1024^3$ grid, with a Taylor-scale Reynolds number ${\rm Re}_\lambda\approx 418$, downloaded from the
Johns Hopkins University turbulence database~\cite{JohnsHopkins,JohnsHopkins2}.
The flow velocity, velocity gradients and second-order gradients were downloaded at the stored time intervals and interpolated linearly to intermediate times.
We integrated the swimmer dynamics using the second-order Adams-Bashforth method with a time step much smaller than the smallest time scale of the swimmer dynamics.

\subsection{DNS of turbulent channel flow}
\label{sec:DNSchannel}
We use in house code for direct numerical simulations of turbulent channel flow~\cite{qiu2019settling} to solve Navier-Stokes equations~(\ref{eq:NSapp}) in a three-dimensional domain surrounded by two infinitely large parallel walls.
A mean pressure gradient is applied in the stream-wise direction to drive the flow. A non-slip boundary condition is applied to the channel wall and periodic boundary conditions are applied to the stream- and span-wise directions.
The resulting flow is characterized by the friction Reynolds number, ${\rm Re}_\tau = hu_\tau / \nu = 180$, where $2h$ is the distance between the walls, $u_\tau=\sqrt{\tau_{\rm wall}/\rho_f}$ is the friction velocity, with $\tau_{\rm wall}$ being the mean shear stress on the wall.
We consider a domain of size $2\pi h \times 2h \times \pi h$ in the stream-wise, wall-normal and span-wise direction, respectively with corresponding mesh size $96 \times 128 \times 96$. The mesh is uniform in the stream- and span-wise directions, and denser close to the walls where the shear is strongest in the wall-normal direction.
Collisions with the channel walls follow the law of reflection: the components of velocity and orientation in the wall-normal direction are reversed, while the components in the stream- and span-wise directions are preserved.
We use a time step $\Delta t u_\tau^2 /\nu = 0.06$, much smaller than the smallest time scale of the swimmer dynamics.
We use a psuedo-spectral method to solve Eqs.~(\ref{eq:NSapp}) in the stream- and span-wise directions, and a second-order finite-difference method in the wall-normal direction. The second-order Adams-Bashforth method is used for time advancement. Similar numerical approach is used in Ref.~\cite{qiu2019settling}.

\section{Robustness of strain avoidance strategy}
\label{sec:robustness}
Fig.~\ref{fig:distributions} shows the distribution of squared strain rate, $\trSS$, similar to Fig.~1({\rm c}), but with a slightly smaller value of $\varepsilon$.
Colored lines show results for the optimal strategy and black lines for tracer particles.
The results illustrate the robustness of the strategy: the swimmer efficiently avoid high-strain regions for large variations of the parameters values.

\section{Comparison to case without thresholds}
\label{sec:sweep_no_threshold}
\begin{figure}[h]
    \begin{overpic}[width=0.4\textwidth]{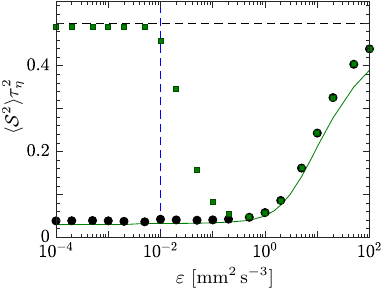}
    \end{overpic}
    \caption{
        \label{fig:sweepSupplemental}
        Average strain $\langle\trSS\rangle$ against the turbulent dissipation rate $\varepsilon$ for the turbulence model with threshold $\Yc=\SI{1}{\per\meter\per\s\squared}$ ($\Box$, green) and without threshold ($\circ$,black). Lines and parameters are same as in Fig.~3 of the main text.
    }
\end{figure}

Fig.~\ref{fig:sweepSupplemental} shows results for the stochastic turbulence model of the average sampled strain $\langle\trSS\rangle$ against energy dissipation rate $\varepsilon$ for swimmers following the optimal policy~(\ref{eq:optimal_policy}) with sensing threshold $\Yc=\SI{1}{\per\meter\per\s\squared}$ (green, square markers) and zero threshold $\Yc=0$ (black, circular markers).
The results show that the threshold does not make much difference for $\varepsilon$ larger than the vertical dashed line, where $\Yc\urms\tauK^3<1$. For smaller $\varepsilon$, where $\Yc\urms\tauK^3>1$, swimmers with a sensing threshold have similar performance as tracer particles (horizontal dashed line), while swimmers without threshold perform almost as good as the simplified model in Eq.~(\ref{eq:eqm_simplified}) (green curve).

%

\end{document}